\documentclass[a4paper, 11pt]{article}
\usepackage{stmaryrd}
\usepackage{}
\usepackage{amsfonts}
\usepackage{dsfont}
\usepackage{mathrsfs}
\usepackage{amssymb}
\usepackage{bbm}
\usepackage{epsfig}
\usepackage{amsmath}
\usepackage{appendix}
\usepackage{color}
\usepackage[english]{babel}

\def\Reals{\mathop{\hbox{\mit I\kern-.2em R}}\nolimits}

\def\Complexes{{\hbox{\mit C\kern-.46em
            \vrule depth 0ex height 1.4ex width .05em\kern.41em}}}

\newtheorem{thm}{Theorem}
\newtheorem{defn}{Definition}

\newtheorem{lem}{Lemma}
\newtheorem{remark}{Remark}
\newtheorem{prop}{Proposition}

\title{\bf  Distributed Optimization: Convergence Conditions\\ from a Dynamical System Perspective\footnote{This work has been supported in part
by the Knut and Alice Wallenberg Foundation, the Swedish Research
Council, and KTH SRA TNG.}}
\date{}

\author{Guodong Shi, Alexandre Proutiere  and  Karl Henrik Johansson\thanks{The authors  are with ACCESS Linnaeus Centre, School of Electrical Engineering,
Royal Institute of Technology, Stockholm 10044, Sweden.
       Email: {\tt\small guodongs@kth.se, alepro@kth.se, kallej@ee.kth.se}}
}

\begin{document}

\maketitle

\begin{abstract}
This paper  explores the fundamental properties of  distributed minimization of  a sum of  functions with
each function only known to one node, and a pre-specified  level of node knowledge and computational capacity.
We define  the optimization information each node receives from its objective function, the neighboring information each node receives from its neighbors, and  the
computational capacity  each node can take advantage of in controlling its state.  It is proven that there exist a neighboring information way and a control law that guarantee global optimal consensus if and only if the solution sets of the local objective functions admit a nonempty intersection set for
fixed strongly connected graphs. Then we show that for any tolerated error, we can find a  control law that guarantees global optimal consensus within this error
 for fixed, bidirectional, and connected graphs under  mild conditions.
For time-varying graphs, we show that optimal consensus can always be achieved  as long as the graph is uniformly jointly strongly connected and the nonempty
intersection condition holds. The results illustrate that nonempty intersection for the local optimal solution sets is a critical
condition for successful  distributed optimization for a large class of algorithms.
\end{abstract}


\section{Introduction}

\subsection{Motivation}
Distributed optimization is on finding a global optimum using local information exchange and cooperative computation over a network. In such problems, there is a global objective function to be
minimized, say, and  each node in the network can only observe part of the objective. The update dynamics is executed through an update equation implemented in each node of the network, based
on the information received from the local objective and the  neighbors.



The literature has not to sufficient extent studied   the real meaning of ``distributed" optimization, or the {\em  level} of
distribution possible for convergence. Some algorithms converge faster than others, while they depend on more information exchange and a more complex
iteration rule. For a precise  study of the level of distribution for optimization methods,  the  way nodes share information, and the computational capacity of each node
 should be specified. Thus, an interesting question arises: fixing the knowledge set and the computational capacity, what is the best performance of any distributed algorithm?
  In this paper, we investigate the fundamental performance limits of distributed algorithms when the constraints on how nodes exchange information and on their computational capacity are fixed. We address these limits from a dynamical system point of view and characterize some fundamental conditions on the global objective function
 for a distributed solution to exist.


\subsection{Related Works}
Distributed optimization is a classical topic in applied mathematics with several excellent textbooks, e.g., \cite{book1,book2,book3}.

Assuming that some estimate of the subgradient for each component of the overall objective function   can be passed over the network from one node to another
via deterministic or randomized iteration, a class of subgradient-based incremental algorithms was investigated  in \cite{solodov, rabbat,nedic01, bjsiam, ram}.
A series of results were established  combining consensus  and subgradient computation. This idea can be traced back
to 1980s to the pioneering work \cite{tsi}.  A subgradient method   for fixed undirected topology was given in
\cite{bj08}. Then in  \cite{nedic09},  convergence bounds for  time-varying
 graphs with various connectivity assumptions  were shown.   This work was then extended to a constrained optimization case in \cite{nedic10}, where each
 agent is assumed to always lie in a particular convex set.
Consensus and optimization were shown to be guaranteed when each node makes a projection onto its own set at each step. Following the ideas of \cite{nedic10},
 a randomized discrete-time  algorithm and a deterministic continuous-time algorithm were presented for optimal consensus in \cite{shirandom} and \cite{shitac}, respectively, where in both cases the goal is to
 form a consensus within the intersection of the optimal solution sets of the local objective functions.  An augmented Lagrangian
algorithm was presented for  constrained optimization with directed gossip communication in \cite{jmf}. An alternative approach was presented in \cite{lu1}, where
the nodes keep their gradient sum equal to zero during the iteration by utilizing gossiping.

 Dynamical system solutions to distributed optimization problem have been  considered for more than fifty years.   The Arrow-Hurwicz-Uzawa flow was shown to converge to the set of saddle points
 for a constrained convex optimization problem \cite{ahu}. In \cite{brockett}, a simple and elegant continuous-time protocol was presented to solve
  linear programming problems. More recently, in  \cite{elia}, a continuous-time solution having second-order   node dynamics was proposed for solving distributed optimization
 problems for fixed bidirectional graphs. In \cite{eben}, a smooth vector field was shown to be able to drive the system trajectory to converge to the saddle point of the Lagrangian of
a convex and constrained optimization problem. In \cite{shitac}, a network of first-order dynamical system was proposed to solve convex intersection computation problems with directed time-varying communication graphs. Besides optimization, a continuous-time interpretation to discrete-time algorithms was  discussed for recursive stochastic algorithms in \cite{ljung}.

Consensus algorithms have been proven to be  useful in the design of distributed optimization methods \cite{nedic09,nedic10,shirandom,shitac,elia,lu1}. Consensus methods  have also been extensively studied for both discrete-time and continuous-time models in the past decade, some references related to the current paper include  \cite{tsi,jad03,julien2,lwang,lin07,ren05,mar,caoming1,mor,nedic08,shi09,shi11}.
\subsection{Main Contribution}
This paper considers the following  distributed optimization model.   The network consists of $N$ nodes with directed communication. Each node $i$  has a convex
objective function $f_i: \mathds{R}^m\rightarrow \mathds{R}$. The goal of the network is to reach consensus meanwhile  minimizing the function $\sum_{i=1}^N f_i$.
At any time $t$, each node $i$ observes  the gradient of $f_i$ at its current state $g_i(t)$ and the neighboring information $n_i(t)$ from its neighbors. The map $n_i(t)$ is zero
when the nodes state is equal to all its neighbors' state.  The evolution of the  nodes' states  is given by a first-order integrator with right-hand side being a control law $\mathcal{J}(n_i,g_i)$ taking feedback from $g_i(t)$ and $n_i(t)$. We assume $\mathcal{J}(n_i,g_i)$  to be injective in $g_i$  when $n_i$ takes value zero.

The main results we obtain are stated as follows:
\begin{itemize}
\item We prove that there exists a  neighboring information rule $n_i$ and a control law $\mathcal{J}$ guaranteeing  global optimal consensus if and only if the intersection of the solution sets of $f_i,i=1,\dots,N$, is nonempty intersection set for
fixed strongly connected graphs.

\item We show that given any $\epsilon>0$, there exists a control law $\mathcal{J}$ that  guarantees global optimal consensus with error no larger than $\epsilon$ for fixed, bidirectional,
and connected graphs under  mild conditions.

\item We show that optimal consensus can always be achieved for time-varying graphs as long as the graph is uniformly jointly strongly connected and the nonempty
intersection condition above holds.
\end{itemize}

We conclude that  the nonempty intersection  of the solution sets of the local objectives seems to be a fundamental condition for distributed optimization.

\subsection{Paper Organization}
In Section~2, some preliminary mathematical concepts and lemmas are introduced. In Section~3, we formulate the considered optimization model, node dynamics, and define
the problem of interest.  Section~4 focuses on fixed graphs. A necessary and sufficient condition is presented for the exact solution of optimal consensus,
and then approximate solutions are investigated as $\epsilon$-optimal consensus. Section~5 is on  time-varying graphs, and we show optimal consensus under uniformly jointly
strongly connected graphs.  Finally, in Section~6 some concluding remarks are
given.

\section{Preliminaries}
In this section, we introduce some  notations and provide preliminary results that will be used in the rest of the paper.

\subsection{Directed Graphs}
A directed graph (digraph) $\mathcal
{G}=(\mathcal {V}, \mathcal {E})$ consists of a finite set
$\mathcal{V}$ of nodes and an arc set
$\mathcal {E}$, where an arc is an ordered pair of
distinct nodes of $\mathcal {V}$ \cite{god}.  An element $(i,j)\in\mathcal {E}$ describes
an arc which leaves $i$ and enters $j$.  A {\it walk} in  $\mathcal
{G}$ is an alternating sequence $\mathcal
{W}:
i_{1}e_{1}i_{2}e_{2}\dots e_{m-1}i_{m}$ of nodes $i_{\kappa}$ and
arcs $e_{\kappa}=(i_{\kappa},i_{\kappa+1})\in\mathcal {E}$ for
$\kappa=1,2,\dots,m-1$. A walk  is called a {\it path}
if the nodes of the walk are distinct, and a path from $i$ to
$j$ is denoted as $i\rightarrow j$. $\mathcal
{G}$ is said to be {\it strongly connected} if it contains path $i\rightarrow j$ and $j\rightarrow i$ for every pair of nodes $i$ and $j$. A digraph $\mathcal {G}$ is called {\it bidirectional}
when for any two nodes $i$ and $j$, $(i,j)\in\mathcal{E}$  if and only if $(j,i)\in\mathcal{E}$. Ignoring the direction of the arcs, the connectivity of a bidirectional digraph is transformed to that of the corresponding undirected graph. A time-varying graph is defined as $\mathcal
{G}_{\sigma(t)}=(\mathcal {V},\mathcal {E}_{\sigma(t)})$ where
$\sigma:[0,+\infty)\rightarrow \mathcal {Q}$ denotes a piecewise constant function,
where $\mathcal {Q}$ is a finite set containing  all possible graphs with node set $\mathcal{V}$. Moreover, the joint graph of $\mathcal
{G}_{\sigma(t)}$ in
time interval $[t_1,t_2)$ with $t_1<t_2\leq +\infty$ is denoted   as
$\mathcal {G}([t_1,t_2))= \cup_{t\in[t_1,t_2)}
\mathcal {G}(t)=(\mathcal {V},\cup_{t\in[t_1,t_2)}\mathcal
{E}_{\sigma(t)})$.

\subsection{Dini Derivatives}
The upper {\it Dini
derivative} of a continuous function $h: (a,b)\to \mathds{R}$ ($-\infty\leq a<b\leq \infty$) at $t$ is defined as
$$
D^+h(t)=\limsup_{s\to 0^+} \frac{h(t+s)-h(t)}{s}.
$$
When $h$ is continuous on $(a,b)$, $h$ is
non-increasing on $(a,b)$ if and only if $ D^+h(t)\leq 0$ for any
$t\in (a,b)$. The next
result is convenient for the calculation of the Dini derivative \cite{dan,lin07}.

\begin{lem}
\label{lemdini}  Let $V_i(t,x): \mathds{R}\times \mathds{R}^d \to \mathds{R}\;(i=1,\dots,n)$ be
$C^1$ and $V(t,x)=\max_{i=1,\dots,n}V_i(t,x)$. If $
\mathcal{I}(t)=\{i\in \{1,2,\dots,n\}\,:\,V(t,x(t))=V_i(t,x(t))\}$
is the set of indices where the maximum is reached at $t$, then
$
D^+V(t,x(t))=\max_{i\in\mathcal{ I}(t)}\dot{V}_i(t,x(t)).
$
\end{lem}

\subsection{Limit Sets}
Consider the following
autonomous system
\begin{equation}
\label{i1} \dot{x}=f(x),
\end{equation}
where $f:\mathds{R}^d\rightarrow \mathds{R}^d$ is a  continuous function. Let $x(t)$ be a solution of
(\ref{i1}) with initial condition $x(t_0)=x^0$. Then $\Omega_0\subset \mathds{R}^d$ is called a {\it positively invariant
set} of (\ref{i1}) if, for any $t_0\in\mathds{R}$ and any $x^0\in\Omega_0$,
we have $x(t)\in\Omega_0$, $t\geq t_0$, along  every solution $x(t)$ of (\ref{i1}).

We call $y$ a  $\omega$-limit point of $x(t)$ if there exists a  sequence $\{t_k\}$ with $\lim_{k\rightarrow \infty}t_k=\infty$ such that
$$
\lim_{k\rightarrow \infty}x(t_k)=y.
$$
The set of all $\omega$-limit points of $x(t)$ is called the  $\omega$-limit set of  $x(t)$,  and is denoted as $\Lambda^+\big(x(t)\big)$.
The following lemma is well-known  \cite{rou}.
\begin{lem}\label{leminvariant}
Let  $x(t)$ be a solution of (\ref{i1}). Then  $\Lambda^+\big(x(t)\big)$ is positively  invariant. Moreover, if $x(t)$ is contained in a compact set,
then $\Lambda^+\big(x(t)\big)\neq \emptyset$.
\end{lem}

\subsection{Convex Analysis}
A set $K\subset \mathds{R}^d$ is said to be {\it convex} if $(1-\lambda)x+\lambda
y\in K$ whenever $x\in K,y\in K$ and $0\leq\lambda \leq1$.
For any set $S\subset \mathds{R}^d$, the intersection of all convex sets
containing $S$ is called the {\it convex hull} of $S$, denoted by
$co(S)$.

Let $K$ be a closed convex subset in $\mathds{R}^d$ and denote
$|x|_K\doteq\inf_{y\in K}| x-y |$ as the distance between $x\in \mathds{R}^d$ and $K$, where $|\cdot|$
is the Euclidean norm.  There is a unique element ${P}_{K}(x)\in K$ satisfying
$|x-{P}_{K}(x)|=|x|_K$ associated to any
$x\in \mathds{R}^d$ \cite{aubin}.  The map ${P}_{K}$ is called the {\it projector} onto $K$. The following lemma holds  \cite{aubin}.
\begin{lem}\label{lemconvex}
(i). $\langle {P}_{K}(x)-x,{P}_{K}(x)-y\rangle\leq 0,\quad \forall y\in
K$.

(ii). $|{P}_{K}(x)-{P}_{K}(y)|\leq|x-y|, x,y\in \mathds{R}^d$.

(iii) $|x|_K^2$ is continuously differentiable at  $x$ with $\nabla |x|_K^2=2\big(x-{P}_{K}(x)\big)$.
\end{lem}

Let  $f: \mathds{R}^d\rightarrow \mathds{R}$ be a real-valued function. We call $f$ a convex function if for any $x,y\in\mathds{R}^d$ and $0\leq\lambda \leq1$,
it holds that $f\big((1-\lambda)x+\lambda y\big)\leq  (1-\lambda)f(x)+\lambda f(y)$. The following lemma states some well-known properties  for convex functions.
\begin{lem}\label{lemfunction} Let $f:\mathds{R}^d\rightarrow \mathds{R}\in C^1$ be a convex function.

(i). $f(x)\geq f(y)+\big\langle x-y, \nabla f(y)\big\rangle$.

(ii). Any local minimum is a global minimum, i.e., $\arg \min f=\big\{z: \nabla f(z)=0 \big\}$.
\end{lem}
\section{Problem Definition}

\subsection{Objective}
Consider a network with node set $\mathcal
{V}=\{1,2,\dots,N\}$ modeled in general as a directed graph $\mathcal{G}=(\mathcal {V}, \mathcal {E})$.  A node $j$ is said to be a {\it neighbor} of $i$ at time $t$ when there is an arc $(j, i)\in \mathcal
{E}$, and we denote $\mathcal{N}_i$  the set of neighbors for node $i$.

 Node $i$ is associated with a cost function $f_i: \mathds{R}^m\rightarrow \mathds{R}, m>0$ which is observed by node $i$ only. The objective for the network  is to  cooperatively solve the
 optimization problem
\begin{equation}\label{1}
     \begin{array}{cl}
       \mathop{\rm minimize}\ & \sum_{i=1}^N f_i(z) \\
       \textrm{subject to} & z\in \mathds{R}^m.
     \end{array}
\end{equation}

We impose the following  assumption on the functions $f_i, i=1,\dots,N$.

\vspace{2mm}

\noindent{\bf A1.} For all $i=1,\dots,N$, we have (i) $f_i\in C^1$; (ii) $f_i$ is a convex function; (iii) $\arg \min f_i\neq \emptyset$.

\vspace{2mm}
Problem (\ref{1}) is equivalent with the following problem:
\begin{equation}\label{16}
     \begin{array}{cl}
       \mathop{\rm minimize}\ & \sum_{i=1}^N f_i(z_i) \\
       \textrm{subject to} & z_i\in \mathds{R}^m  \\
       & z_1=\dots=z_N.
     \end{array}
\end{equation}
From (\ref{16}) we see that  consensus algorithms are a natural mean  for  solving the optimization problem  (\ref{1}).

\subsection{Information Flow}
The state of node $i$ at time $t$ is denoted as $x_i(t)\in \mathds{R}^m$. We define the  information flow for node  $i$ as  follows.

\begin{itemize}
\item The local optimization information $g_i(t)$  node $i$ receives from its objective $f_i$ at time $t$  is  the gradient of $f_i$ at its current state, i.e.,
\begin{align}
g_i(t)\doteq \nabla f_i\big(x_i(t)\big).
\end{align}

\item The neighboring information $n_i(t)$ node $i$ receives from its neighbors at time $t$ is
\begin{align}
n_i(t)\doteq  \hbar_i\big(x_i(t), x_j(t): j \in
\mathcal{N}_i \big),
\end{align}
where $\hbar_i: \mathds{R}^m\times \mathds{R}^{m|\mathcal{N}_i|}\rightarrow \mathds{R}^l$ is a continuous function, $|\mathcal{N}_i|$ denotes  the number of elements
in $\mathcal{N}_i$,  and $l$ is a given integer indicating the dimension of the neighboring information.

\end{itemize}

Let $\hbar= \hbar_1 \otimes \dots \otimes \hbar_N:  \mathds{R}^{m(1+|\mathcal{N}_1|)} \times \dots \times \mathds{R}^{m(1+|\mathcal{N}_N|)}  \rightarrow \mathds{R}^{Nl}$ denote the direct sum of $\hbar_i, i=1,\dots,N$. Then  $\hbar$ represents the rule of  all  neighboring information flow over the whole network. We impose the following assumption.

\vspace{2mm}

\noindent {\bf A2.} $\hbar\in \mathscr{R} \doteq \Big\{  h_1 \otimes \dots \otimes h_N$: \  $h_i$: $ \mathds{R}^{m(1+|\mathcal{N}_i|)} \mapsto\mathds{R}^{l}$ and $h_i\equiv0$ within the local consensus manifold $\big\{x_i=x_{j}: j \in
\mathcal{N}_i \big\}$ for all $i\in\mathcal{V}\Big\}$.


\begin{remark}
 Assumption A2 is to say that the neighboring information a node receives from its neighbors becomes trivial
 when the node is in the same state as all its neighbors. This is a quite natural assumption in the literature
 on distributed averaging and optimization algorithms  \cite{jad03,mor,saber04,nedic09,nedic10}.
\end{remark}

\subsection{Computational Capacity}
We adopt a  dynamical system model to define the way nodes update their respective states. The evolution of the nodes' states is restricted  to be a  first-order integrator:
\begin{equation}\label{2}
\dot{x}_i=u_i, \quad i=1,\dots,N,
\end{equation}
where  the right-hand side $u_i$ is
interpreted as a control input and the control law is characterized as
\begin{align}\label{5}
u_i= \mathcal {J} \big(n_i,g_i\big),\ i=1,\dots,N
\end{align}
with $\mathcal{J}:
\mathds{R}^l\times \mathds{R}^m\rightarrow \mathds{R}^m$.

For the control law $\mathcal{J}$, we impose the following assumption.

\vspace{2mm}

\noindent {\bf A3.}
 $\mathcal{J} \in \mathscr{C}\doteq \big\{\mathcal{F}(\cdot,\cdot)\in C^0:\
\mathds{R}^l\times \mathds{R}^m\rightarrow \mathds{R}^m,\ \mathcal{F}(0,\cdot)\ \mbox{is injective}\big\}$.

\begin{remark}
Assumption A3  indicates that the control law applied in each node should have the same structure, irrespectively of   individual  local optimization information or neighboring information.
Note that our network model is homogeneous because one cannot tell the difference from one node to another.
We assume that the control law  $\mathcal{J}(0,\cdot)$ is injective, so each node takes different response to different gradient information
on the local consensus manifold. Again,  Assumption A3 is widely applied in the literature \cite{jad03,mor,saber04,nedic09,nedic10}.
\end{remark}

\subsection{Problem}
Let $x(t)=(x_1^T(t),\dots,x_N^T(t))^T\in \mathds{R}^{mN}$ be the trajectory of system (\ref{2}) with control law (\ref{5}) for initial condition $x^0=x(t_0)$.
Denote  $F(z)=\sum_{i=1}^N f_i(z)$. We introduce the following definition.

\begin{defn} Global {\it optimal consensus}  of  (\ref{2})--(\ref{5}) is achieved if for all $x^0\in \mathds{R}^{mN}$, we have
\begin{equation}\label{3}
\limsup_{t\rightarrow +\infty} F\big(x_i(t)\big)= \min_{z\in \mathds{R}^m} F(z)
\end{equation}
and
\begin{equation}\label{4}
\lim_{t\rightarrow +\infty} \big |x_i(t)-x_j(t)\big|=0,\quad i,j=1,\dots,N.
\end{equation}
\end{defn}

The problem considered in this paper is to characterize conditions on the control law $\mathcal{J}$ under which global optimal consensus is achieved. In Section~4 this is done for fixed graphs and in Section~5 for time-varying graphs.
\section{Fixed Graphs}
In this section, we consider  the possibility of solving optimal consensus using control law (\ref{5}) under fixed communication graphs. We first discuss
whether exact optimal consensus can be reached for  directed graphs. Then we show the existence of an approximate solution for optimal consensus over bidirectional
graphs.
\subsection{Exact Solution}
 We make an assumption on the solution set of $F=\sum_{i=1}^N f_i$.

\noindent{\bf A4. } $\arg \min F(z)\neq\emptyset$ is a bounded set.

The main result on the existence of a control law solving optimal consensus is stated as follows.

\begin{thm}\label{thm1}Assume that A1 and A4 hold. Let the communication graph $\mathcal{G}$ be fixed and  strongly connected.
There exist  a  neighboring information rule $\hbar \in \mathscr{R}$ and a control law $\mathcal{J}\in \mathscr{C}$ such that global optimal consensus is achieved if and only if
\begin{align}\label{intersection}
\bigcap_{i=1}^N \arg \min  f_i(z)\neq \emptyset.
\end{align}
\end{thm}

\begin{remark}
According to Theorem \ref{thm1}, the optimal solution sets of $f_i$, $i=1,\dots,N$, having nonempty intersection is a critical condition for the existence of a
control law (\ref{5}) that solves the optimal consensus problem.
Condition (\ref{intersection}) is obviously a strong constraint
which in general does not hold. Therefore, basically Theorem \ref{thm1} suggests that exact solution of optimal consensus is seldom possible  for the given model.
\end{remark}

\begin{remark}
It follows from the proof below that the necessity statement of Theorem~\ref{thm1} relies  only on the fact that the limit set of an autonomous system is invariant. It is straightforward to verify that for a discrete-time autonomous dynamical system defined by
\begin{align}
y_{k+1}=f(y_k)
\end{align}
with $f$ a continuous function, its limit set is  invariant. Therefore, if
we consider a model with   discrete-time update  as
\begin{align}\label{203}
x_i(k+1)=x_i(k)+u_i(k)
\end{align}
with \begin{align}\label{204}
u_i(k)=\mathcal{J}\big(n_i(k),g_i(k)\big),
\end{align}
where $n_i$, $g_i$, and $\mathcal{J}$ agree with the  definitions above, the necessity statement of Theorem \ref{thm1} still holds. However, the sufficiency
statement of Theorem \ref{thm1} may in general not hold for discrete-time updates since even for the centralized optimization problem,
there is not always an algorithm with constant step size which can solve the problem exactly, cf., \cite{bertsekas}.

\end{remark}
 \begin{remark}
In \cite{nedic09}, a discrete-time algorithm was provided for solving (\ref{1}), where the structure of the nodes' update is the sum of
  a consensus term averaging the neighbors' states,
 and a subgradient term of the local objective function with a fixed step size. It is  easy to see that the algorithm in \cite{nedic09} can be rewritten as (\ref{203}) and (\ref{204}) as long as the graph is fixed and the step size is constant. All the properties we impose on the information flow and update dynamics are kept. Convergence bounds
 were established for the case with constant step size in \cite{nedic09}.  Theorem \ref{thm1} shows  that proposing a convergence bound is in general the best we
 can do for algorithms like the one developed in \cite{nedic09}, and the result also explains why a time-varying step size may be necessary in distributed optimization algorithms, as
 in \cite{nedic10}.

\end{remark}

In the rest of this subsection, we first give the proof of the necessity claim of  Theorem \ref{thm1}, and then we present a simple proof for the sufficiency part with bidirectional
graphs.  The sufficiency part of Theorem \ref{thm1}  in fact follows from the upcoming conclusion, Theorem \ref{thm4}, which does not rely on Assumption  A4.
\subsubsection{Necessity}
We now prove the necessity statement in Theorem \ref{thm1} by a contradiction argument. Suppose $\bigcap_{i=1}^N \arg \min  f_i(z)= \emptyset$ and
there exists a distributed control in the form of (\ref{5}), say $\mathcal {J}_0 \big(n_i,g_i\big)$,
under which global optimal consensus is reached for certain neighboring information flow $n_i$ satisfying Assumption A2. Let $x(t)$ be  a trajectory of system (\ref{2}) with control
$\mathcal {J}_0 \big(n_i,g_i\big)$ and $\Lambda^+(x(t))$ be its  $\omega$-limit set. The definition of optimal consensus leads to that $x(t)$ converges to the bounded set
$\Big( \arg \min F(z)\Big)^N\bigcap \mathcal {M}$,
where $\Big( \arg \min F(z)\Big)^N$ denotes the $N$'th power set of $ \arg \min F(z)$ and  $\mathcal{M}$ denotes the consensus manifold,  defined by
\begin{align}
\mathcal{M}\doteq \big\{x=(x_1^T\dots x_N^T)^T: \ x_1=\dots=x_N;\ x_i\in \mathds{R}^m, i=1,\dots,N\big\}.
\end{align}
Therefore, each trajectory $x(t)$ is contained in a compact set.

Based on Lemma \ref{leminvariant}, we conclude that  $\Lambda^+(x(t))\neq \emptyset$ and
\begin{align}\label{6}
\Lambda^+(x(t))\subseteq \Big( \arg \min F(z)\Big)^N\bigcap \mathcal {M},
\end{align}
Moreover, $\Lambda^+(x(t))$ is  positively invariant since system (\ref{2}) is autonomous
under control $\mathcal {J}_0 \big(n_i,g_i\big)$ when the communication graph is fixed. This is to say, any trajectory  of system (\ref{2})
 under control $\mathcal {J}_0 \big(n_i,g_i\big)$ must stay within $\Lambda^+(x(t))$ for any initial value in $\Lambda^+(x(t))$.

Now we take $y\in\Lambda^+(x(t))$. Then we have $y\in  \Big( \arg \min F(z)\Big)^N\bigcap \mathcal {M}$ according to (\ref{6}),
and thus  $y=(z_\ast^T \dots z_\ast^T)^T$ for some  $z_\ast\in \arg \min F(z)$. With Assumption A1, the convexity of the $f_i$'s implies that
\begin{align}
\arg \min F(z)=\big\{z\in\mathds{R}^m:\ \sum_{i=1}^N \nabla f_i(z)=0 \big\}.
\end{align}
On the other hand, we have
$$
\bigcap_{i=1}^N \arg \min  f_i(z)= \bigcap_{i=1}^N \big\{z\in\mathds{R}^m:\ \nabla f_i(z)=0 \big\}= \emptyset.
$$
Therefore, there exists two indices $i_1,i_2\in \{1,\dots,N\}$ with $i_1\neq i_2$ such that
\begin{align}
\nabla f_{i_1}(z_\ast)\neq \nabla f_{i_2}(z_\ast).
\end{align}

Consider the solution of (\ref{2})  under control $\mathcal {J}_0 \big(n_i,g_i\big)$ for initial time $t_0$ and initial value $y$.
The fact that $y$ belongs to the consensus manifold guarantees
\begin{align}
n_{i_1}(t_0)=n_{i_2}(t_0)=0.
\end{align}
With Assumption A4, we have
\begin{align}
\mathcal {J}_0\big(n_{i_1}(t_0), g_{i_1}(t_0)\big)= \mathcal {J}_0\big(0, \nabla f_{i_1}(z_\ast)\big) \neq
\mathcal {J}_0\big(0, \nabla f_{i_2}(z_\ast)\big)= \mathcal {J}_0\big(n_{i_2}(t_0), g_{i_2}(t_0)\big).
\end{align}
This implies $\dot{x}_{i_1}(t_0)\neq \dot{x}_{i_2}(t_0)$. As a result, there exists a constant $\varepsilon>0$
such that $x_{i_1}(t)\neq x_{i_2}(t)$ for $t\in (t_0,t_0+\varepsilon)$. In other word, the trajectory will leave the set
$$
\Big( \arg \min F(z)\Big)^N\bigcap \mathcal {M}
$$
for $(t_0,t_0+\varepsilon)$, and therefore will also leave the set $\Lambda^+(x(t))$. This contradicts the fact that  $\Lambda^+(x(t))$ is positively invariant.
The necessity part of Theorem \ref{thm1} has been proved.
\subsubsection{Sufficiency: Bidirectional Case}
We now provide an alternative proof of sufficiency for bidirectional graphs, which is based on  some geometrical intuition of the vector field. Note that compared to the proof of Theorem \ref{thm4} on directed graphs, this proof uses completely different arguments which indeed cannot be applied to directed graphs. Therefore, we believe the proof given in the following is interesting at its own right, because it reveals some fundamental difference between directed and bidirectional graphs.

Let $a_{ij}>0$ be a constant marking the weight of arc $(j,i)$. We will show  that the particular neighboring information flow
$$
n_i=\sum\limits_{j \in
\mathcal{N}_i}a_{ij}\big(x_j-x_i\big)
$$
and control law
\begin{align}\label{7}
\mathcal{J}_\star(n_i,g_i)=n_i-g_i=\sum\limits_{j \in
\mathcal{N}_i}a_{ij}\big(x_j-x_i\big)-\nabla f_i\big(x_i\big)
\end{align}
ensure global optimal consensus for system (\ref{2}). Note that (\ref{7}) is indeed a continuous-time version of the algorithm proposed in \cite{nedic09}.

We suppose $\mathcal{G}$ is bidirectional. In this case, we have $a_{ij}=a_{ji}$ for all $i$ and $j$, and we use unordered pair $\{i,j\}$  to denote the edge between node $i$ and $j$.

Noticing  that
\begin{align}
\mathcal{J}_\star(n_i,g_i)=\sum\limits_{j \in
\mathcal{N}_i}a_{ij}\big(x_j-x_i\big)-\nabla f_i\big(x_i\big)=-\nabla_{x_i} \Big(\frac{1}{2}\sum\limits_{j \in
\mathcal{N}_i}a_{ij}\big|x_j-x_i\big|^2+f_i(x_i)\Big),
\end{align}
we have that $(\ref{7})$ indeed solves the following convex problem
\begin{align}\label{a1}
     \begin{array}{cl}
       \mathop{\rm minimize}\ & F_{\mathcal{G}}(x)\doteq\sum_{i=1}^N f_i(x_i)+ \frac{1}{2}\sum_{\{j,i\}\in\mathcal{E}}a_{ij}\big|x_j-x_i\big|^2 \\
       \textrm{subject to} & x_i\in \mathds{R}^m,\ i=1,\dots,N.
     \end{array}
\end{align}
We establish the following lemma  relating the   solution sets of problems  (\ref{1}) and (\ref{a1}).
\begin{lem}\label{lem2}
Suppose $\bigcap_{i=1}^N \arg \min  f_i(z)\neq \emptyset$. Suppose also the communication graph $\mathcal{G}$ is fixed, bidirectional, and  connected. Then we have
\begin{align}
\arg \min  F_{\mathcal{G}}(x)=\Big( \bigcap_{i=1}^N \arg \min  f_i(z)\Big)^N\bigcap \mathcal {M}=\Big( \arg \min F(z)\Big)^N\bigcap \mathcal {M}.
\end{align}
\end{lem}
{\it Proof.} When $\bigcap_{i=1}^N \arg \min  f_i(z)\neq \emptyset$, it is straightforward to see that
$$
\arg \min F(z)=  \bigcap_{i=1}^N \arg \min  f_i(z).
$$

Now take $x_\ast=(p_\ast^T \dots p_\ast^T)^T\in \Big( \bigcap_{i=1}^N \arg \min  f_i(z)\Big)^N\bigcap \mathcal {M}$, where $p_\ast\in  \bigcap_{i=1}^N \arg \min  f_i(z)$.
First we have $x_\ast\in \arg \min_x \sum_{i=1}^N f_i(x_i) $. Second we have $x_\ast\in \arg \min_x \frac{1}{2}\sum_{\{j,i\}\in\mathcal{E}}a_{ij}\big|x_j-x_i\big|^2$.
Therefore, we conclude that $x_\ast\in \arg \min  F_{\mathcal{G}}(x) $. This gives
\begin{align}\label{11}
\arg \min  F_{\mathcal{G}}(x) \supseteq\Big( \bigcap_{i=1}^N \arg \min  f_i(z)\Big)^N\bigcap \mathcal {M}.
\end{align}

On the other hand, convexity gives
\begin{align}
\arg \min  F_{\mathcal{G}}(x)=\bigg\{x:\ -(L\otimes I_m) x= \Big( \big(\nabla f_1(x_1)\big)^T \dots \big(\nabla f_N(x_N)\big)^T \Big)^T\bigg\},
\end{align}
where $\otimes$ represents the Kronecker product, $I_m$ is the identity matrix in $\mathds{R}^m$,  and  $L=D-A$ is the Laplacian of the graph $\mathcal{G}$ with $A=[a_{ij}]$ and $D={\rm diag}(d_1,\dots,d_N)$,
where $d_i=\sum_{j=1}^n a_{ij}$. Noticing that $$
(\mathbf{1}_N^T\otimes I_m) (L\otimes I_m) =\mathbf{1}_N^TL\otimes I_m=0,
$$
where $\mathbf{1}_N=(1\dots 1)^T \in \mathds{R}^N$, we have
\begin{align}\label{8}
\Big(\mathbf{1}_N^T \otimes I_m\Big) \Big( \big(\nabla f_1(x_1)\big)^T \dots \big(\nabla f_N(x_N)\big)^T \Big)^T=\sum_{i=1}^N \nabla f_i(x_i)=0
\end{align}
for any $x\in\arg \min  F_{\mathcal{G}}(x) $.

Now take $x^\ast=(q_1^T \dots q_N^T)^T\in\arg \min  F_{\mathcal{G}}(x) $. Suppose there exist two indices $i_\ast$ and $j_\ast$ such that
$$
\nabla f_{i_\ast}(q_{i_\ast})\neq \nabla f_{j_\ast}(q_{j_\ast}).
$$
Then at least one of $\nabla f_{i_\ast}(q_{i_\ast})$ and $\nabla f_{j_\ast}(q_{j_\ast})$ must be nonzero. Taking $\hat{p}\in  \bigcap_{i=1}^N \arg \min  f_i(z)$, we have
$$
\sum_{i=1}^N f_i(q_i)>\sum_{i=1}^Nf_i(\hat{p})
$$
because for $x=(x_1^T \dots x_N^T)^T\in\arg \min  \sum_{i=1}^N f_i(x_i)$, we have $\nabla f_i(x_i)=0, i=1,\dots,N$.
Consequently, for $w_\ast=(\hat{p}^T \dots \hat{p}^T)^T$, we have
$$
F_{\mathcal{G}}(x^\ast)>F_{\mathcal{G}}(w_\ast),
$$
which is impossible according to the definition of $x^\ast$ so that such $i_\ast$ and $j_\ast$ cannot exist. In light of (\ref{8}), this  immediately implies
$$
\nabla f_i(q_i)=0,\ i=1,\dots,N,
$$
or equivalently
\begin{align}\label{10}
q_i\in \arg \min f_i(z),\ i=1,\dots,N
\end{align}
 for all $x^\ast=(q_1^T\dots q_N^T)^T\in\arg \min  F_{\mathcal{G}}(x)$.

 Therefore, we conclude from (\ref{10}) that
 $$
 \sum_{i=1}^N f_i(q_i)=\sum_{i=1}^Nf_i(p_\ast),
 $$
 and this implies
 $$
\sum_{\{j,i\}\in\mathcal{E}}a_{ij}\big|q_j-q_i\big|^2=0
 $$
 as long as $x^\ast=(q_1^T \dots q_N^T)^T\in\arg \min  F_{\mathcal{G}}(x)$. The connectivity of the communication graph thus further guarantees  that $q_1=\dots=q_N$,
so we have proved that
 $
 x^\ast\in \Big( \bigcap_{i=1}^N \arg \min  f_i(z)\Big)^N\bigcap \mathcal {M}.$ Consequently, we obtain
\begin{align}\label{12}
\arg \min  F_{\mathcal{G}}(x) \subseteq\Big( \bigcap_{i=1}^N \arg \min  f_i(z)\Big)^N\bigcap \mathcal {M}.
\end{align}

The desired lemma holds from (\ref{11}) and (\ref{12}). \hfill$\square$

Now since $F_{\mathcal{G}}(x)$ is a convex function and we have $\dot{x}=\nabla F_{\mathcal{G}}(x)$ for system (\ref{2}) with control (\ref{7}), we conclude that
$$
\lim_{t\rightarrow \infty}{\rm dist}\big(x(t),{\arg \min F_{\mathcal{G}}(x) }\big)=0.
$$
Lemma \ref{lem2} ensures
$$
\lim_{t\rightarrow \infty}{\rm dist}\bigg(x(t),\Big( \bigcap_{i=1}^N \arg \min  f_i(z)\Big)^N\bigcap \mathcal {M}\bigg)=0
$$
if $\mathcal{G}$ is bidirectional and connected.
Equivalently, global optimal consensus is reached.

\begin{remark}
We see from the proof above that the construction of $F_{\mathcal{G}}(x)$    is critical because the convergence argument is based on the fact that
the gradient of $F_{\mathcal{G}}(x)$ is consistent with the communication graph.
It can be easily verified that finding such a function  is in general impossible for directed graphs.
\end{remark}

\subsection{Approximate Solution}
Theorem \ref{thm1} indicates that optimal consensus is impossible no matter how the control law $\mathcal{J}$ is chosen from $\mathscr{C}$
as long as the nonempty intersection condition (\ref{intersection}) is not fulfilled. In this subsection, we discuss the approximate
solution  of the optimal consensus problem in the absence of (\ref{intersection}).
 We introduce the  following definition.

\begin{defn}
 Global {\it $\epsilon$-optimal consensus}  is achieved if for all $x^0\in \mathds{R}^{mN}$, we have
\begin{equation}\label{3a}
\limsup_{t\rightarrow +\infty} F\big(x_i(t)\big)\leq \min_{z\in \mathds{R}^m} F(z)+\epsilon
\end{equation}
and
\begin{equation}\label{4a}
\lim_{t\rightarrow +\infty} \big |x_i(t)-x_j(t)\big|\leq \epsilon,\quad i,j=1,\dots,N.
\end{equation}
\end{defn}

Denoting $F_{\mathcal{G}}(x;K)=\sum_{i=1}^N f_i(x_i)+ \frac{K}{2}\sum_{\{j,i\}\in\mathcal{E}}a_{ij}\big|x_j-x_i\big|^2$, we impose the following assumption.

\noindent{\bf A5. } (i) $\arg \min F(z)\neq\emptyset$;
(ii) $\arg \min F_{\mathcal{G}}(x;K)\neq\emptyset$ for all $K\geq 0$;
(iii) $\bigcup_{K\geq 0} \arg \min F_{\mathcal{G}}(x;K)$ is bounded.

\vspace{2mm}

For $\epsilon$-optimal consensus, we present the following result.
\begin{thm}\label{thm2}
Assume that A1 and A5  hold. Let the communication graph $\mathcal{G}$ be fixed, bidirectional, and connected. Then for any $\epsilon>0$,
there exist  a  neighboring information rule $\hbar \in \mathscr{R}$ and a control law $\mathcal{J}\in \mathscr{C}$ such that global $\epsilon$-optimal consensus is achieved.
\end{thm}
{\it Proof. } Again, let $a_{ij}>0$ be any constant marking the weight of arc $(j,i)$ and $a_{ij}=a_{ji}$ for all $(i,j)\in \mathcal{E}$. Fix $\epsilon$. We will show  that under neighboring information flow
$$
n_i=\sum\limits_{j \in
\mathcal{N}_i}a_{ij}\big(x_j-x_i\big),
$$
 there exists a   constant $K_\epsilon>0$ such that  the control law
\begin{align} \label{50}
u_i=\mathcal{J}_{K_\epsilon}(n_i,g_i)\doteq K_\epsilon n_i-g_i
\end{align}
guarantees global $\epsilon$-optimal consensus.

 It is straightforward to see that
\begin{align}
\mathcal{J}_{K}(n_i,g_i)=K\sum\limits_{j \in
\mathcal{N}_i}a_{ij}\big(x_j-x_i\big)-\nabla f_i\big(x_i\big)=-\nabla_{x_i} \Big(\frac{K}{2}\sum\limits_{j \in
\mathcal{N}_i}a_{ij}\big|x_j-x_i\big|^2+f_i(x_i)\Big).
\end{align}
System (\ref{2}) with control law
$u_i=\mathcal{J}_K(n_i,g_i)$ can be written into the following compact form
\begin{align}\label{90}
\dot{x}=-\nabla F_{\mathcal{G}}(x;K),\ \ x=(x_1^T \dots x_N^T)^T\in \mathds{R}^{mN}.
\end{align}
Then the convexity of $F_{\mathcal{G}}(x;K)$ ensures that control law $\mathcal{J}_{K}(n_i,g_i)$ asymptotically solves  the  convex optimization problem
\begin{align}\label{ka1}
     \begin{array}{cl}
       \mathop{\rm minimize}\ & F_{\mathcal{G}}(x;K)=\sum_{i=1}^N f_i(x_i)+ \frac{K}{2}\sum_{\{j,i\}\in\mathcal{E}}a_{ij}\big|x_j-x_i\big|^2 \\
       \textrm{subject to} & x_i\in \mathds{R}^m,\  i=1,\dots,N.
     \end{array}
\end{align}

Convexity gives
\begin{align}\label{92}
\arg \min  F_{\mathcal{G}}(x;K)=\bigg\{x:\ -K(L\otimes I_m) x= \Big( \big(\nabla f_1(x_1)\big)^T \dots \big(\nabla f_N(x_N)\big)^T \Big)^T\bigg\}.
\end{align}
Under Assumptions A1 and A5, we have that
\begin{align}
L_0\doteq \sup \Big\{ \big|\nabla \tilde{F}(x)\big|:\ x\in \bigcup_{K\geq0} \arg \min  F_{\mathcal{G}}(x;K)\Big\}
\end{align}
is a finite number, where $\tilde{F}(x)=\sum_{i=1}^N f_i(x_i)$. We also define
\begin{align}
D_0\doteq  \sup \Big\{ \big|z_\ast-x_i\big|:\ i=1,\dots,N,\  x\in \bigcup_{K\geq 0} \arg \min  F_{\mathcal{G}}(x;K)\Big\},
\end{align}
where $z_\ast\in \arg \min F$ is an arbitrarily chosen point.

Let $p=(p_1^T \dots p_N^T)^T \in \arg \min  F_{\mathcal{G}}(x;K)$ with $p_i\in \mathds{R}^m, i=1, \dots,N$. Since the graph is bidirectional and connected, we can sort the eigenvalues of the Laplacian $L\otimes I_m$ as
$$
0=\lambda_1=\dots=\lambda_{m} <\lambda_{m+1}\leq \dots \leq\lambda_{mN}.
$$
Let $l_1\dots,l_{mN}$ be the orthonormal basis of $\mathds{R}^{mN}$ formed by the right eigenvectors of $L\otimes I_m$,
where  $l_1,\dots,l_m$ are eigenvectors corresponding to the zero eigenvalue. Suppose $p=\sum_{k=1}^{mN}c_k l_{k}$ with $c_k\in\mathds{R}, k=1,\dots,mN$.

According to (\ref{92}),  we have
\begin{align}
\Big|K(L\otimes I_m)p\Big|^2= K^2\Big|\sum_{k=m+1}^{mN} c_k \lambda_k l_k\Big|^2=K^2\sum_{k=m+1}^{mN} c_k^2 \lambda_k^2 \leq L_0^2,
\end{align}
which yields
\begin{align}\label{94}
\sum_{k=m+1}^{mN} c_k^2 \leq \Big(\frac{L_0}{K\lambda_2^\ast}\Big)^2,
\end{align}
where $\lambda_2^\ast>0$ denotes the second smallest  eigenvalue of $L$.

Now recall that
\begin{align}
\mathcal{M}\doteq \big\{x=(x_1^T\dots x_N^T)^T: \ x_1=\dots=x_N;\ x_i\in \mathds{R}^m, i=1,\dots,N\big\}.
\end{align}
is the consensus manifold. Noticing that $\mathcal{M}={\rm span} \{l_1,\dots, l_m\}$, we conclude from (\ref{94}) that
\begin{align}\label{100}
\sum_{k=m+1}^{mN} c_k^2 =\Big| \sum_{k=m+1}^{mN} c_k l_k \Big|^2=| p |_{\mathcal{M}}^2=\sum_{i=1}^N \Big|p_i-\frac{\sum_{i=1}^{N} p_i}{N}\Big|^2\leq \Big(\frac{L_0}{K\lambda_2^\ast}\Big)^2.
\end{align}
The last equality in (\ref{100}) is due to the fact that $\mathbf{1}_N \otimes\Big(  \frac{\sum_{i=1}^{N} p_i}{N}\Big)$ is the projection of $p$ on to $\mathcal{M}$. Thus, for any $\varsigma >0$, there is $K_1(\varsigma)>0$ such that when $K\geq K_1(\varsigma)$,
\begin{align}\label{101}
\Big|p_i-p_{\rm ave}\Big|\leq \varsigma,\ i=1,\dots,N
\end{align}
and
\begin{align}
|F(p_i)-F(p_{\rm ave})\Big|\leq \varsigma,\ i=1,\dots,N,
\end{align}
where $p_{\rm ave}=\frac{\sum_{i=1}^N p_i}{N}$.

On the other hand, with (\ref{92}), we have
\begin{align}\label{102}
\sum_{i=1}^N \nabla f_i(p_i)=\sum_{i=1}^N \nabla f_i(p_{\rm ave}+\hat{p}_i)=0,
\end{align}
where  $\hat{p}_i=p_i-p_{\rm ave}$. Now according to (\ref{101}) and (\ref{102}),  since  each $f_i\in C^1$, for any $\varsigma >0$,
there is $K_2(\varsigma)>0$ such that when $K\geq K_2(\varsigma)$,
\begin{align}
\Big|\sum_{i=1}^N \nabla f_i(p_{\rm ave}) \Big|\leq \frac{\varsigma}{D_0}.
\end{align}
This implies
\begin{align}
F(p_{\rm ave})\leq F(z_\ast)+|z_\ast-p_{\rm ave}|\times \Big|\sum_{i=1}^N \nabla f_i(p_{\rm ave}) \Big|\leq F(z_\ast)+\varsigma.
\end{align}

Therefore, for any $\epsilon>0$, we can take $K_0=\max\{K_1(\epsilon/2), K_2(\epsilon/2) \}$. Then when $K\geq K_0$, we have
\begin{align}
|p_i-p_j|\leq \epsilon;\ \ F(p_i)\leq \min_z F(z)+\epsilon
\end{align}
for all $i$ and $j$. Now that  $F_{\mathcal{G}}(x;K)$ is a convex function and observing (\ref{90}), every limit point of  system (\ref{2}) with control law  $\mathcal{J}_K(n_i,g_i)$
is contained in the set $\arg \min F_{\mathcal{G}}(x;K)$. Noting that $p$ is arbitrarily chosen from $\arg \min F_{\mathcal{G}}(x;K)$, $\epsilon$-optimal consensus is achieved as long
as we choose $K_\epsilon\geq K_0$.  This completes the proof. \hfill$\square$

\begin{remark}
Theorem \ref{thm2} can be compared to the results given in \cite{solodov}, where a discrete-time incremental algorithm with constant step size was shown to be able to reach an $\epsilon$-approximate
solution of  (\ref{1}). Incremental algorithms relies on global iteration along each local objective function alternatively \cite{solodov,ram,bjsiam}. They are therefore fundamentally different
with the model we discuss.
\end{remark}
\begin{remark}
 For the discrete-time algorithm proposed in  \cite{nedic09}, a bound of the convergence error was  expressed explicitly as a function of
the fixed step size. However, this bound will  not vanish as the  fixed step size tends to zero or infinity \cite{nedic09}. Note that the parameter $K$ in
the control law $\mathcal{J}_{K}(n_i,g_i)$ can be viewed as a step size. As shown in Theorem \ref{thm2},
 the convergence error vanishes as $K$ tends to infinity, which is essentially  different with the discrete-time case in \cite{nedic09}.
\end{remark}

From Theorems \ref{thm1} and \ref{thm2}, we conclude that even though  without the nonempty intersection condition (\ref{intersection}), it is impossible to reach exact optimal consensus via  control law of  the form of (\ref{5}), it is still possible to find a control law  that guarantees  approximate optimal consensus with arbitrary accuracy.

\subsection{Discussion: Global vs. Local}
A fundamental question in distributed optimization is whether global optimization can be obtained by neighboring information flow and cooperative computation. We have the following observation.

\begin{itemize}
\item  Note that in this paper, to determine  a proper $K$ in (\ref{50}) for a given $\epsilon$ relies on knowledge of  the structure of the network, and the information of all
 $f_i, i=1,\dots,N$. Finding a proper control law  for $\epsilon$-optimal consensus requires thus {\em global} knowledge of the  network.
  Apparently also the nonempty intersection condition in Theorem \ref{thm1} is a  {\em global} constraint.

\item Incremental algorithms with constant step size have been shown to be able to reach $\epsilon$-optimal solution for any error bound $\epsilon$ as long as the step size is sufficiently small, e.g.,  \cite{solodov,nedic01,bjsiam}. In an incremental algorithm, iteration is carried out by only one node alternatively on each local objective function, which is
    is equivalent to the fact that the $N$ nodes perform the iteration, but any node can access the states of all other nodes. Therefore, it means that the underlying graph is indeed complete, which is certainly a {\em global} constraint.

    \item One can also use time-varying step size. In \cite{nedic10}, it was shown that global optimization can be achieved by a algorithm combining consensus algorithm and subgradient computation with a time-varying step size. However, this time-varying step size must be applied to all nodes homogeneously, which makes it a {\em global} parameter.
    \end{itemize}

    From the above observations, we can conclude that in general for distributed optimization methods,    some { global}
  information (or constraint) is somehow inevitable  to guarantee  a global (exact or $\epsilon$-approximate) convergence. This reveals  some fundamental limit of distributed information collection and algorithm design.
\subsection{Assumption Feasibility}
This subsection discusses  the feasibility of Assumptions A4 and A5 and shows that
some  mild conditions are enough to  ensure A4 and A5.

\begin{prop}
Let  A1  hold. If $\tilde{F}(x)=\sum_{i=1}^N f_i(x_i)$ is coercive, i.e., $\tilde{F}(x)\rightarrow \infty$ as long as $|x|\rightarrow \infty$,  then A4 and A5 hold.
\end{prop}
{\it Proof.} Assume that A1 holds. 

 a). Since $\tilde{F}(x)=\sum_{i=1}^N f_i(x_i)$ is coercive, it follows straightforwardly that $F(z)=\sum_{i=1}^Nf_i(z)$ is also coercive.
As a result, $\arg \min F(z)\neq \emptyset$ is a bounded set. Thus, A4 and A5.(i) hold.
%

b).  Observing that $\frac{K}{2}\sum_{\{j,i\}\in\mathcal{E}}a_{ij}\big|x_j-x_i\big|^2\geq 0$ for all $x=(x_1^T \dots x_N^T)^T\in \mathds{R}^{mN}$
and that $\tilde{F}(x)=\sum_{i=1}^N f_i(x_i)$ is coercive,
we obtain that  $\arg \min F_{\mathcal{G}}(x;K)\neq\emptyset$ for all $K\geq 0$. Thus,  A5.(ii) holds.

c). Based on a), we can denote $F_\ast=\min_z F(z)=F(z_\ast)$.  Since $\sum_{i=1}^Nf_i(x_i)$ is coercive,  there exists a constant $M(F_\ast)>0$ such that $\sum_{i=1}^N f_i(x_i)> F_\ast$ for all
$|x|>M$. This implies
\begin{align}
F_{\mathcal{G}}(x;K)> F_{\mathcal{G}}(\mathbf{1}_N\otimes z_\ast;K)=F_\ast
\end{align}
for all $|x|> M$. That is to say, the global minimum of $F_{\mathcal{G}}(x;K)$ is reached within the set $\{|x|\leq M\}$ for all $K>0$. Therefore, we have
\begin{align}
\bigcup_{K\geq 0} \arg \min F_{\mathcal{G}}(x;K)\subseteq  \big\{|x|\leq M\big\}.
\end{align}
This proves A5.(iii). \hfill$\square$


Next, we propose another case when A4 and A5 hold.
\begin{prop}
Let A1 hold.  Suppose each $\arg \min f_i$ is bounded and the argument space for each $f_i$ is $\mathds{R}$, i.e., $m=1$. Then A4 and A5 holds.
\end{prop}
{\it Proof.}  Assume that A1 holds. 

a). Let $x_i^\ast \in \arg \min f_i$.  Denote $y_\ast=\min \{x_1^\ast,\dots,x_N^\ast\}$. Then for any $i=1,\dots,N$, we have
\begin{align}
0\geq f_i(x_i^\ast)-f_i(y_\ast)\geq (x_i^\ast -y_\ast)\nabla f_i(y_\ast)
\end{align}
according to inequality (i) of Lemma \ref{lemfunction}. This immediately yields $\nabla f_i(y_\ast)\leq 0$ for all $i=1,\dots,N$.

Thus, for any $y<y_\ast$, we have
\begin{align}
F(y)-F(y_\ast)\geq (y -y_\ast)\nabla F(y_\ast)=\sum_{i=1}^N (y -y_\ast) \nabla f_i(y_\ast)\geq 0,
\end{align}
which implies $F(y)\geq F(y_\ast)$ for all  $y<y_\ast$.

A symmetric analysis leads to that $F(y)\geq F(y^\ast)$ for all  $y>y^\ast$ with $y^\ast=\max \{x_1^\ast,\dots,x_N^\ast\}$. Therefore, we obtain
$ F(y)\geq \min\{F(y_\ast), F(y^\ast)\}$ for all $y\neq [y_\ast, y^\ast]$. This implies that a global minimum is reached within
the interval $[y_\ast, y^\ast]={\rm co}\{x_1^\ast,\dots,x_N^\ast\}$  and A5.(i) thus follows.

If $\arg \min f_i$ is bounded for $i=1,\dots,N$, there exist $b_i\leq d_i, i=1,\dots, N$ such that $\arg \min f_i=[b_i, d_i]$. Define $b_\ast =\min\{b_1,\dots,b_N\}$ and
$d^\ast= \max\{d_1,\dots,d_N\}$. Following a similar argument we have $\arg \min F \subseteq [b_\ast, d^\ast]$. Thus A4 holds.

b).  Introduce the following cube in $\mathds{R}^N$:
$$
\mathcal{C}_\ast^\eta\doteq \Big\{x=(x_1^T \dots x_N^T)^T: \ x_i \in [y_\ast-\eta, y^\ast+\eta],i=1,\dots,N\Big\},
$$
where $\eta>0$ is a given constant.

\vspace{1mm}

\noindent {\it Claim.} For any $K\geq 0$, $\mathcal{C}_\ast^\eta$ is an invariant set of system (\ref{2}) under control law  $\mathcal{J}_K(n_i,g_i)$.

\vspace{1mm}

Define $\Psi(x(t))=\max_{i\in\mathcal{V}} x_i(t)$. Then based on Lemma \ref{lemdini}, we have
\begin{align}
D^+\Psi(x(t))&=\max_{i\in \mathcal{I}_0(t)} \frac{d}{dt}x_i(t)\nonumber\\
&=\max_{i\in \mathcal{I}_0(t)}  \sum\limits_{j \in
\mathcal{N}_i}a_{ij}\big(x_j-x_i\big)-\nabla f_i\big(x_i\big) \nonumber\\
&\leq  \max_{i\in \mathcal{I}_0(t)} \Big[-\nabla f_i\big(x_i\big) \Big],
\end{align}
where $\mathcal{I}_0(t)$ denotes the index set which contains all the nodes reaching the maximum for $\Psi(x(t))$.

Since
\begin{align}
0\geq f_i(x_i^\ast)-f_i(y_\ast+\eta)\geq (x_i^\ast -y_\ast-\eta)\nabla f_i(y_\ast+\eta),\ i=1,\dots,N
\end{align}
we have $\nabla f_i( y^\ast+\eta)\geq 0$ for all $i=1,\dots,N$. As a result, we obtain
\begin{align}
D^+\Psi(x(t))\Big|_{\Psi(x(t))=y^\ast+\eta}\leq 0,
\end{align}
which implies $\Psi(x(t))\leq y^\ast+\eta$ for all $t\geq t_0$ under initial condition $\Psi(x(t_0))\leq y^\ast+\eta$. Similar analysis ensures that
$\min_{i\in\mathcal{V}} x_i(t)\geq y^\ast-\eta$ for all $t\geq t_0$ as long as $\min_{i\in\mathcal{V}} x_i(t_0)\geq y^\ast-\eta$. This proves the claim.

\vspace{2mm}

Note that every trajectory of system (\ref{2}) under control law  $\mathcal{J}_K(n_i,g_i)$ asymptotically solves (\ref{ka1}). This immediately leads to
that $ F_{\mathcal{G}}(x;K)$ reaches its minimum within   $\mathcal{C}_\ast^\eta$ for any $K\geq 0$ since $\mathcal{C}_\ast^\eta$ is an invariant set.
Then  A5.(ii) holds straightforwardly.

c). Since  $\arg \min f_i$ is bounded for $i=1,\dots,N$, there exist $b_i\leq d_i, i=1,\dots, N$ such that $\arg \min f_i=[b_i, d_i]$. Define $b_\ast =\min\{b_1,\dots,b_N\}$ and
$d^\ast= \max\{d_1,\dots,d_N\}$. We will prove the conclusion by showing $\arg \min F_{\mathcal{G}}(x;K) \subseteq \mathcal{C}_\ast$ for all $K\geq 0$, where
$$
\mathcal{C}_\ast\doteq \Big\{x=(x_1^T \dots x_N^T)^T: \ x_i \in [b_\ast, d^\ast],i=1,\dots,N\Big\}.
$$

Let $z=(z_1 \dots,z_N)^T\in \arg \min F_{\mathcal{G}}(x;K)$. First we show $\max\{z_1,\dots,z_N\} \leq d^\ast$ by a contradiction argument. Suppose $\max\{z_1,\dots,z_N\} > d^\ast$.

Now let $i_1,\dots, i_k$ be the nodes reaching the maximum state, i.e.,  $z_{i_1}=\dots=z_{i_k}=\max\{z_1,\dots,z_N\}$. There will be two cases.
\begin{itemize}
\item Let $k=N$. We have $z_1=\dots=z_N=y$ in this case. Then for all $i$ and $x_i^\ast \in \arg \min f_i$, we have
\begin{align}
0>f_i(x_i^\ast)-f_i(y)\geq (x_i^\ast -y) \nabla f_i(y)
\end{align} which yields $ \nabla f_i(y)>0, i=1,\dots,N$ since $y>d^\ast$. This immediately leads to
\begin{align}
F_{\mathcal{G}}(z;K)=F(y)>\min F \geq \min F_{\mathcal{G}}(z;K),
\end{align}
which contradicts the fact that $z\in \arg \min F_{\mathcal{G}}(x;K)$.
\item Let $k<N$. Then we denote  $s_\ast=\max\big \{ z_i: i\notin\{i_1,\dots,i_k\}, i=1,\dots,N \big\}$, which is actually the second largest value in $\{z_1,\dots,z_N\}$.
We define a new point $\hat{z}=(\hat{z}_1 \dots,\hat{z}_N)^T$ by $\hat{z}_i=z_i, i\notin\{i_1,\dots,i_k\}$ and
	\begin{align}
\hat{z}_i=\begin{cases}
		d^\ast, & \mbox{if $s_\ast<d^\ast$}\\
	s_\ast, & \mbox{otherwise}
	\end{cases}
\end{align}
for $i\in \{i_1,\dots,i_k\}$. Then it is easy to obtain that $F_{\mathcal{G}}(z;K)>F_{\mathcal{G}}(\hat{z};K)$, which again contradicts the choice of $z$.
\end{itemize}
Therefore, we have proved that $\max\{z_1,\dots,z_N\} \leq d^\ast$. Based on a  symmetric analysis we also have $\min\{z_1,\dots,z_N\} \geq b_\ast$. Therefore, we obtain
$\arg \min F_{\mathcal{G}}(x;K) \subseteq \mathcal{C}_\ast$ for all $K\geq 0$ and A5.(iii) follows. \hfill$\square$
\section{Time-varying Graphs}
Now we consider time-varying graphs. The communication in the multi-agent network is modeled as $\mathcal {G}_{\sigma(t)}=(\mathcal {V},\mathcal
{E}_{\sigma(t)})$ with
$\sigma:[0,+\infty)\rightarrow \mathcal {Q}$ being a piecewise constant function,
where $\mathcal {Q}$ is a finite set indicating all possible graphs. In this case the  neighbor set for each node is time-varying,
and we let $\mathcal{N}_i(\sigma(t))$ represent the set of agent $i$'s neighbors at time $t$. As usual in the literature \cite{jad03,lin07,shi09},
an assumption is given to how fast $\mathcal {G}_{\sigma(t)}$ can vary.

\vspace{2mm}
\noindent {\bf A6.} {\it (Dwell Time)} There is a lower bound $\tau_D>0$ between two consecutive
switching time instants of $\sigma(t)$.
\vspace{2mm}

\vspace{2mm}
We have the following definition.
\begin{defn} (i) $\mathcal
{G}_{\sigma(t)}$ is said to be {\it uniformly jointly strongly connected}  if there exists a constant $T>0$
such that $\mathcal {G}([t,t+T))$ is strongly connected for any $t\geq0$.

(ii) $\mathcal
{G}_{\sigma(t)}$ is said to be {\it uniformly jointly quasi-strongly connected}  if there exists a constant $T>0$
such that $\mathcal {G}([t,t+T))$ has a spanning tree for any $t\geq0$.
\end{defn}

 \vspace{2mm}

With time-varying graphs,
\begin{align}
n_i(t)\doteq  \hbar_i\big(x_i(t), x_j(t): j \in
\mathcal{N}_i(\sigma(t)) \big).
\end{align}
where $\hbar_i: \mathds{R}^m\times \mathds{R}^{m|\mathcal{N}_i(\sigma(t))|}\rightarrow \mathds{R}^l$ is now  piecewise defined. As a result, assumption A2 is transformed to the following piecewise version.

\noindent {\bf A7.} $\hbar\in \mathscr{R}_\ast \doteq \Big\{  h_1 \otimes \dots \otimes h_N$:  $h_i$ maps $ \mathds{R}^{m(1+|\mathcal{N}_i(\sigma(t))|)}$ to $\mathds{R}^{l}$ on each time interval when $\sigma(t)$ is constant,  and $h_i\equiv0$ within the  time-varying local consensus manifold $\big\{x_i=x_{j}: j \in
\mathcal{N}_i (\sigma(t))\big\}$ for all $i\in\mathcal{V}\Big\}$.

For optimal consensus with time-varying graphs, we present the following result.
\begin{thm}\label{thm3} Suppose   A1 and A6 hold  and $\mathcal
{G}_{\sigma(t)}$ is uniformly jointly strongly connected. Suppose  $\bigcap_{i=1}^N \arg \min  f_i\neq \emptyset$  contains at least one interior point.
Then there exist  a  neighboring information rule $\hbar \in \mathscr{R}_\ast$ and a control law $\mathcal{J}\in \mathscr{C}$ such that global optimal consensus is achieved and
 \begin{align}\label{17}
 \lim_{t\rightarrow\infty}x_i(t)=x_\ast.
 \end{align}
for some $x_\ast\in \bigcap_{i=1}^N \arg \min  f_i$.
\end{thm}

Note that (\ref{17}) is indeed a stronger conclusion than our definition of optimal consensus as Theorem \ref{thm3} guarantees that all the node states converge to a common point
in the global solution set of $F(z)$. We will see from the proof of Theorem \ref{thm3} that this state convergence highly relies on the existence of an interior point
of $\bigcap_{i=1}^N \arg \min  f_i$. In the absence of such an interior point condition, it turns out that optimal consensus still stands. We present another theorem stating the fact.

\begin{thm}\label{thm4} Suppose   A1 and A6 hold  and $\mathcal
{G}_{\sigma(t)}$ is uniformly jointly strongly connected. Suppose also $\bigcap_{i=1}^N \arg \min  f_i\neq \emptyset$.
Then there exist  a  neighboring information rule $\hbar \in \mathscr{R}_\ast$ and a control law $\mathcal{J}\in \mathscr{C}$ such that global optimal consensus is achieved.
\end{thm}

The proofs of Theorems \ref{thm3} and \ref{thm4} rely on the following  neighboring information flow
\begin{align}\label{201}
n_i=\sum\limits_{j \in
\mathcal{N}_i(\sigma(t))}a_{ij}(t)\big(x_j-x_i\big),
\end{align}
where  $a_{ij}(t)>0$ is any weight function associated with arc $(j,i)$. The resulting control law is
\begin{align}
\mathcal{J}_\star(n_i,g_i)=n_i-g_i.
\end{align}
An assumption is made on  each $a_{ij}(t),i,j=1,2,...,N$.

\vspace{2mm}
\noindent{\bf A8.} {\it (Weights Rule)}  (i) Each $a_{ij}(t)$ is piece-wise continuous and $a_{ij}(t)\geq0$ for all $i$ and $j$.

\noindent (ii). There are $a^\ast>0$ and
$a_\ast>0$ such that $
 a_\ast\leq a_{ij}(t)\leq a^\ast,\quad
 t\in \mathds{R}^+.$

\subsection{Preliminary  Lemmas}
We establish three useful lemmas in this subsection.

Suppose $\bigcap_{i=1}^N \arg \min  f_i\neq \emptyset$ and take $z_\ast\in \bigcap_{i=1}^N \arg \min  f_i$. We define
\begin{align}
V_i(t)=\big|x_i(t)-z_\ast\big|^2,\ i=1,\dots,N,
\end{align}
and
\begin{align}
V(t)=\max_{i=1,\dots,N} V_i(t).
\end{align}
The following lemma holds with the proof in Appendix A.1.
\begin{lem}\label{lemmono} Let  A1 and A8 hold.  Suppose $\bigcap_{i=1}^N \arg \min  f_i\neq \emptyset$.
Then along any trajectory of system  (\ref{2}) with neighboring information (\ref{201})  and control law $\mathcal{J}_\star(n_i,g_i)$, we have $D^+V(t)\leq 0$ for all $t\in \mathds{R}^+$.
\end{lem}

A direct consequence of Lemma \ref{lemmono} is that when $\bigcap_{i=1}^N \arg \min  f_i\neq \emptyset$, we have
\begin{align}
\lim_{t\rightarrow \infty} V(t)=d_\ast^2
\end{align}
for some $d_\ast\geq 0$ along any trajectory of system  (\ref{2}) with control law $\mathcal{J}_\star(n_i,g_i)$.
However, it is still unclear whether $V_i(t)$ converges or not. We establish another lemma indicating that with proper connectivity condition for the communication graph, all $V_i(t)$'s
have the same limit $d_\ast^2$. The proof can be found in Appendix A.2.

\begin{lem}\label{lemlimit}
Let A1,  A6, and A8 hold. Suppose $\bigcap_{i=1}^N \arg \min  f_i\neq \emptyset$ and $\mathcal
{G}_{\sigma(t)}$ is uniformly jointly strongly  connected. Then along any trajectory of system  (\ref{2}) with neighboring information (\ref{201})  and control law $\mathcal{J}_\star(n_i,g_i)$,
we have $\lim_{t\rightarrow \infty} V_i(t)=d_\ast^2 $ for all $i$.
\end{lem}

The next lemma shows that each node will reach its own optimum along the trajectories of system  (\ref{2}) under
control law $\mathcal{J}_\star(n_i,g_i)$. The proof is in Appendix A.3.

\begin{lem}\label{lemnodeoptimum}
Let A1,  A6, and A8 hold. Suppose $\bigcap_{i=1}^N \arg \min  f_i\neq \emptyset$ and $\mathcal
{G}_{\sigma(t)}$ is uniformly jointly strongly  connected. Then along any trajectory of system  (\ref{2}) with control law $\mathcal{J}_\star(n_i,g_i)$,
we have $\limsup_{t\rightarrow \infty} \big| x_i(t)\big|_{\arg\min f_i}=0$ for all $i$.
\end{lem}
\subsection{Proof of Theorem \ref{thm3}}
The proof of Theorem \ref{thm3} relies on the following lemma.
\begin{lem}\label{lemunique} Let $z_1,\dots,z_{m+1}\in\mathds{R}^m$ and $d_1,\dots,d_{m+1}\in\mathds{R}^+$.
Suppose there exist solutions to equations (with variable $y$)
\begin{equation}\label{23}
	\begin{cases}
		|y-z_1|^2 =d_1;\\
		\ \ \ \ \ \  \vdots\\
|y-z_{m+1}|^2 =d_{m+1}.
	\end{cases}
\end{equation}
Then the solution  is unique if ${\rm rank}\big(z_2-z_1, \dots, z_{m+1}-z_1\big)=m$.
\end{lem}
{\it Proof.} Take $j>1$ and let $y$ be a solution to the equations. Noticing that
$$
\langle y-z_1,y-z_1\rangle=d_1; \quad \langle y-z_j,y-z_j\rangle=d_j
$$
we obtain
\begin{align}
\langle y,z_j-z_1\rangle= \frac{1}{2}\Big(d_1-d_j+|z_j|^2-|z_1|^2\Big), \ j=2,\dots,m+1.
\end{align}
The desired conclusion follows immediately. \hfill$\square$

We now prove Theorem \ref{thm3}. Let $r_\star=(r_1^T \dots r_N^T)^T$ be a limit point of a trajectory of system  (\ref{2}) with control law $\mathcal{J}_\star(n_i,g_i)$.

We first show consensus. Based on Lemma \ref{lemlimit}, we have $\lim_{t\rightarrow \infty} V_i(t)=d_\ast$ for all $z_\ast\in\bigcap_{i=1}^N \arg \min  f_i$. This is to say,
$|r_i-z_\ast|=d_\ast$ for all $i$ and $z_\ast\in\bigcap_{i=1}^N \arg \min  f_i$.
Since $\bigcap_{i=1}^N \arg \min  f_i\neq \emptyset$ contains at least one interior point, it is obvious to see that we can find $z_1,\dots,z_{m+1}\in\bigcap_{i=1}^N \arg \min  f_i$ with  ${\rm rank}\big(z_2-z_1, \dots, z_{m+1}-z_1\big)=m$ and $d_1,\dots,d_{m+1}\in\mathds{R}^+$, such that each $r_i, i=1,\dots,N$ is a solution of equations (\ref{23}). Then based on Lemma \ref{lemunique}, we conclude that $r_1=\dots=r_N$. Next, with Lemma \ref{lemnodeoptimum}, we have  $| r_i|_{\arg\min f_i}=0$. This implies that $r_1=\dots=r_N\in\bigcap_{i=1}^N \arg \min  f_i$, i.e., optimal consensus is achieved.

We turn to state convergence. We only need to  show that $r_\star$ is unique along any trajectory of system  (\ref{2}) with neighboring information (\ref{201})  and control law $\mathcal{J}_\star(n_i,g_i)$. Now suppose
$r_\star^1=\mathbf{1}_N\otimes r^1$ and $r_\star^2=\mathbf{1}_N\otimes r^2$ are two different limit points with $r^1\neq r^2 \in \bigcap_{i=1}^N \arg \min  f_i$. According to the definition of a
limit point, we have that for any $\varepsilon>0$, there exists a time instant $t_\varepsilon$ such that $|x_i(t_\varepsilon)-r^1|\leq \varepsilon$ for all $i$.
Note that Lemma \ref{lemmono} indicates that the disc $B(r^1,\varepsilon)=\{y: |y-r^1|\leq \varepsilon\}$ is an invariant set for initial time $t_\varepsilon$.
While taking $\varepsilon={|r^1-r^2|}/{4}$, we see that $r^2\notin  B(r^1,|r^1-r^2|/{4})$. Thus, $r^2$ cannot be a limit point.

Now since the limit point is unique, we denote it as  $\mathbf{1}_N\otimes x_\ast$ with $x_\ast\in \bigcap_{i=1}^N \arg \min  f_i$.
Then we have  $\lim_{t\rightarrow\infty}x_i(t)=x_\ast$ for all $i=1,\dots,N$. This completes the proof.

\subsection{Proof of Theorem \ref{thm4}}
In this subsection, we prove Theorem \ref{thm4}. We need the following lemma on robust consensus, which can be found in \cite{shicdc}.

\begin{lem}\label{lemrobust}
Consider a network with node set $\mathcal{V}=\{1,\dots,N\}$ with time-varying communication graph $\mathcal{G}_{\sigma(t)}$. Let the dynamics of node $i$ be
\begin{align}
\dot{x}_i=\sum\limits_{j \in
\mathcal{N}_i(\sigma(t))}a_{ij}(t)\big(x_j-x_i\big)+w_i(t),
\end{align}
where $w_i(t)$ is a piecewise continuous  function. Suppose A6 and A8 hold and $\mathcal{G}_{\sigma(t)}$ is uniformly jointly quasi-strongly connected.  Then we have  \begin{equation}
\lim_{t\rightarrow +\infty} \big |x_i(t)-x_j(t)\big|=0,\quad i,j=1,\dots,N
\end{equation}
if $\lim_{t\rightarrow \infty}w_i(t)=0$ for all $i$.
\end{lem}

Lemma \ref{lemnodeoptimum} indicates that $\limsup_{t\rightarrow \infty} \big| x_i(t)\big|_{\arg\min f_i}=0$ for all $i$, which yields
\begin{align}
\lim_{t\rightarrow \infty}\nabla f_i\big(x_i(t)\big)=0
\end{align}
for all $i$ according to Assumption A1. Then the consensus part in the definition of optimal consensus follows immediately from Lemma \ref{lemrobust}.
Again by  Lemma \ref{lemnodeoptimum},
we further conclude that
$\limsup_{t\rightarrow \infty} {\rm dist}\big( x_i(t), \bigcap_{i=1}^N \arg \min  f_i\big)=0$. The desired conclusion thus follows.

\section{Conclusions}
Various algorithms have been proposed in the literature for the distributed
minimization of $\sum_{i=1}^N f_i$ with
$f_i$ only known to node $i$. This paper explored some fundamental properties for  distributed methods
 given a certain level of node knowledge,  computational capacity, and  information flow. It was proven that there exists  a control law   that ensures global optimal consensus if and only if  $\arg \min f_i, i=1,\dots,N$,
 admit a nonempty intersection set for fixed strongly connected graphs. We also showed  that for any error bound, we can find a control law which guarantees global optimal consensus within this bound
 for fixed, bidirectional, and connected graphs under some mild conditions such as  that $f_i$ is coercive for some $i$.
For time-varying graphs, it was proven  that optimal consensus can always be achieved  as long as the graph is uniformly jointly strongly connected
and the nonempty
intersection condition holds. It was then concluded that  nonempty intersection for the local optimal solution sets is a critical
condition for distributed optimization using consensus processing.

More challenges lie in exploring the corresponding  limit of performance for high-order schemes, the optimal structure of the underlying communication graph for distributed optimization,  and the fundamental  communication complexity required for global convergence.

\section*{Acknowledgment }
The authors would like to thank Prof. Angelia Nedi\'{c} for helpful discussions and for pointing out relevant literature.

\section*{Appendix}

\subsection*{\bf A.1\ \  Proof of Lemma \ref{lemmono} }
 Based on Lemma \ref{lemdini}, we have
\begin{align}\label{20}
D^+V(t)&=\max_{i\in \mathcal{I}(t)} \frac{d}{dt}V_i(t)\nonumber\\
&=\max_{i\in \mathcal{I}(t)} 2\Big\langle x_i(t)-z_\ast, \sum\limits_{j \in
\mathcal{N}_i(\sigma(t))}a_{ij}(t)\big(x_j-x_i\big)-\nabla f_i\big(x_i\big)\Big\rangle,
\end{align}
where $\mathcal{I}(t)$ denotes the index set which contains all the nodes reaching the maximum for $V(t)$.

Let $m\in\mathcal{I}(t)$. Denote
$$
Z_t=\big\{z:\ |z-z_\ast|\leq \sqrt{V(t)} \big\}
$$
as the disk centered at $z_\ast$ with radius $\sqrt{V(t)}$. Take $y=x_m(t)+(x_m(t)-z_\ast)$. Then from some simple Euclidean geometry it is obvious to see that $P_{Z_t}(y)=x_m(t)$,
where $P_{Z_t}$ is the {\it projector} onto $Z_t$. Thus, for all $j\in\mathcal{N}_{m}(\sigma(t))$, we obtain
\begin{align}\label{18}
\big\langle x_m(t)-z_\ast,x_j(t)-x_m(t)\big\rangle&=\big\langle y-x_m(t),x_j(t)-x_m(t)\big\rangle\nonumber\\
&=\big\langle y-P_{Z_t}(y),x_j(t)-P_{Z_t}(y)\big\rangle\nonumber\\
&\leq 0
\end{align}
according to inequality (i) in Lemma \ref{lemconvex} since $x_j(t)\in Z_t$. On the other hand, based on inequality (i) in Lemma \ref{lemfunction}, we also have
\begin{align}\label{19}
\big\langle x_m(t)-z_\ast,-\nabla f_m\big(x_m(t)\big)\big\rangle\leq f_m(z_\ast)-f_m\big(x_m(t)\big) \leq 0
\end{align}
in light of the definition of $z_\ast$.

With (\ref{20}), (\ref{18}) and (\ref{19}), we conclude that
\begin{align}
D^+ V(t) =\max_{i\in \mathcal{I}(t)} 2\big\langle x_i(t)-z_\ast, \sum\limits_{j \in
\mathcal{N}_i(\sigma(t))}a_{ij}(t)\big(x_j-x_i\big)-\nabla f_i\big(x_i\big)\big\rangle\leq  0,
\end{align}
which completes the proof. \hfill $\square$

\subsection*{\bf A.2\ \  Proof of Lemma \ref{lemlimit} }
In order to prove the desired conclusion, we just need to show $\liminf_{t\rightarrow \infty} V_i(t)=d_\ast^2$ for all $i$.

With Lemma \ref{lemmono}, we conclude that $\forall\varepsilon>0, \exists M(\varepsilon)>0$, s.t.,
\begin{align}\label{24}
\sqrt{V_i(t)}\leq d_\ast+\varepsilon
\end{align}
for all $i$ and $t\geq M$.

\vspace{2mm}
{\it Claim.} For all $t\geq M$ and all $i,j\in\mathcal{V}$, we have
\begin{align}\label{25}
\big\langle x_i(t)-z_\ast,x_j(t)-x_i(t) \big\rangle\leq-V_i(t)+(d_\ast+\varepsilon)\sqrt{V_i(t)}.
\end{align}

If $x_i(t)=z_\ast$ (\ref{25}) follows trivially from (\ref{24}). Otherwise we take $y_\ast= z_\ast+ (d_\ast+\varepsilon)\frac{x_i(t)-z_\ast}{|x_i(t)-z_\ast|}$ and $B_t=\big\{z: |z-z_\ast|\leq d_\ast+\varepsilon\big \}$. Here $B_t$ is the disk centered at $z_\ast$ with radius $d_\ast+\varepsilon$, and $y_\ast$ is a point within the boundary of $B_t$ and falls the same line with $z_\ast$ and $x_{i_0}(t)$. Take also $q_\ast=y_\ast+x_i(t)-z_\ast$. Then we have
\begin{align}
\big\langle x_i(t)-z_\ast,x_j(t)-y_\ast \big\rangle&=\big\langle q_\ast-y_\ast,x_j(t)-y_\ast \big\rangle\nonumber\\
&=\big\langle q_\ast-P_{B_t}(q_\ast),x_j(t)-P_{B_t}(q_\ast) \big\rangle\nonumber\\
&\leq 0
\end{align}
according to inequality (i) in Lemma \ref{lemconvex}, which leads to
\begin{align}
\big\langle x_i(t)-z_\ast,x_j(t)-x_i(t) \big\rangle&=\big\langle x_i(t)-z_\ast,x_j(t)-y_\ast \big\rangle+\big\langle x_i(t)-z_\ast,y_\ast-x_i(t) \big\rangle\nonumber\\
&\leq\big\langle x_i(t)-z_\ast,y_\ast-x_i(t)\big\rangle\nonumber\\
&=-V_i(t)+(d_\ast+\varepsilon)\sqrt{V_i(t)}.
\end{align}
This proves the claim.

\vspace{2mm}

Now suppose there exists $i_0\in\mathcal{V}$ with $\liminf_{t\rightarrow \infty} V_i(t)=\theta_{i_0}^2<d_\ast^2$. Then we can find a time sequence $\{t_k\}_1^\infty$ with $\lim_{k\rightarrow \infty}t_k =\infty$ such that
\begin{align}\label{28}
\sqrt{V_{i_0}(t_k)}\leq \frac{\theta_{i_0}+d_\ast}{2}.
\end{align}
We divide the rest of the proof into three steps.

\noindent {\it Step 1.}
Take $t_{k_0}>M$. We bound $V_{i_0}(t)$ in this step.

With the weights rule A8, (\ref{25}) and inequality (i) in Lemma \ref{lemfunction}, we see that
\begin{align}\label{45}
\frac{d}{dt} V_{i_0}(t)&=2\Big \langle x_{i_0}(t)-z_\ast, \sum\limits_{j \in
\mathcal{N}_{i_0}(\sigma(t))}a_{i_0j}(t)\big(x_j-x_{i_0}\big)-\nabla f_{i_0}\big(x_{i_0}(t)\big) \Big\rangle\nonumber\\
&\leq 2\sum\limits_{j \in
\mathcal{N}_{i_0}(\sigma(t))}a_{i_0j}(t)  \Big \langle x_{i_0}(t)-z_\ast,x_j(t)-x_{i_0}(t) \Big\rangle+f_{i_0}\big(z_\ast\big)-f_{i_0}\big(x_{i_0}(t)\big)\nonumber\\
&\leq 2(N-1)a^\ast\Big(-V_{i_0}(t)+(d_\ast+\varepsilon)\sqrt{V_{i_0}(t)}\Big),
\end{align}
for all $t\geq t_{k_0}$, which implies
\begin{align}\label{26}
\frac{d}{dt}\sqrt{V_{i_0}(t)} \leq -(N-1)a^\ast\Big(\sqrt{V_{i_0}(t)}-(d_\ast+\varepsilon)\Big),\ \ t\geq t_{k_0}.
\end{align}

In light of Gr\"{o}nwall's inequality, (\ref{28}) and (\ref{26}) yield
\begin{align}\label{30}
\sqrt{V_{i_0}(t)} &\leq e^{-(N-1)^2a^\ast T_D}\sqrt{V_{i_0}(t_{k_0})}+\Big(1-e^{-(N-1)^2a^\ast T_D}\Big)(d_\ast+\varepsilon)\nonumber\\
&\leq \frac{e^{-(N-1)^2a^\ast T_D}}{2} \theta_{i_0}+\Big(1-\frac{e^{-(N-1)^2a^\ast T_D}}{2}\Big)(d_\ast+\varepsilon)\nonumber\\
&\doteq \Lambda_\ast.
\end{align}
for all $t\in[t_{k_0}, t_{k_0}+(N-1)T_D]$ with $T_D=T+\tau_D$, where $T$ comes from the definition of uniformly jointly strongly connected graphs and $\tau_D$ represents the dwell time.

\noindent {\it Step 2.} Since the graph is uniformly jointly strongly connected, we can find an instant $\hat{t}\in[t_{k_0},t_{k_0}+T]$ and another node $i_1\in\mathcal{V}$ such that $(i_0,i_1)\in\mathcal{G}_{\sigma(t)}$ for $t\in[\hat{t}, \hat{t}+\tau_D]$. In this step, we continue to bound $V_{i_1}(t)$.

 Similar to (\ref{25}), for all $t\geq M$ and all $i,j\in\mathcal{V}$, we also have
\begin{align}\label{29}
\big\langle x_i(t)-z_\ast,x_j(t)-x_i(t) \big\rangle\leq-\sqrt{V_i(t)}\Big(\sqrt{V_i(t)}-\sqrt{V_j(t)}\Big)
\end{align}
when $V_j(t)\leq V_i(t)$. Then based on (\ref{25}), (\ref{30}), and (\ref{29}), we obtain
\begin{align}\label{31}
\frac{d}{dt} V_{i_1}(t)
&\leq 2\sum\limits_{j \in
\mathcal{N}_{i_1}(\sigma(t))}a_{i_1j}(t)  \Big \langle x_{i_1}(t)-z_\ast,x_j(t)-x_{i_1}(t) \Big\rangle\nonumber\\
&= 2\sum\limits_{j \in
\mathcal{N}_{i_1}(\sigma(t))\setminus \{i_0\}}a_{i_1j}(t)  \Big \langle x_{i_1}(t)-z_\ast,x_j(t)-x_{i_1}(t) \Big\rangle+2a_{i_1i_0}(t)  \Big \langle x_{i_1}(t)-z_\ast,x_{i_0}(t)-x_{i_1}(t) \Big\rangle\nonumber\\
&\leq 2(N-2)a^\ast\Big(-V_{i_1}(t)+(d_\ast+\varepsilon)\sqrt{V_{i_1}(t)}\Big)-2a_\ast\sqrt{V_{i_1}(t)}\Big(\sqrt{V_{i_1}(t)}-\sqrt{V_{i_0}(t)}\Big)\nonumber\\
&\leq- 2\Big((N-2)a^\ast+a_\ast\Big)V_{i_1}(t) +2\sqrt{V_{i_1}(t)} \Big((N-2)a^\ast(d_\ast+\varepsilon)+\Lambda_\ast a_\ast\Big)
\end{align}
for $t\in[\hat{t},\hat{t}+\tau_D]$, where without loss of generality we assume $V_{i_1}(t)\geq V_{i_0}(t)$ during all $t\in[\hat{t},\hat{t}+\tau_D]$.

Then (\ref{31}) gives
\begin{align}
\frac{d}{dt} \sqrt{V_{i_1}(t)}
&\leq- \Big((N-2)a^\ast+a_\ast\Big)\sqrt{V_{i_1}(t)} + \Big((N-2)a^\ast(d_\ast+\varepsilon)+\Lambda_\ast a_\ast\Big), t\in[\hat{t},\hat{t}+\tau_D]
\end{align}
which yields
\begin{align}
\sqrt{V_{i_1}(\hat{t}+\tau_D)}&\leq e^{- \big((N-2)a^\ast+a_\ast\big)\tau_D}(d_\ast+\varepsilon)+\Big(1-e^{- \big((N-2)a^\ast+a_\ast\big)\tau_D}\Big)\frac{(N-2)a^\ast(d_\ast+\varepsilon)+\Lambda_\ast a_\ast}{(N-2)a^\ast+a_\ast}\nonumber\\
&=\frac{ a_\ast\big(1-e^{- ((N-2)a^\ast+a_\ast)\tau_D}\big)}{(N-2)a^\ast+a_\ast}\times\frac{e^{-(N-1)^2a^\ast T_D}}{2} \theta_{i_0}\nonumber\\
&\ \ \ \ \ \ \ \ \ \ \ +\Big(1-\frac{ a_\ast\big(1-e^{- ((N-2)a^\ast+a_\ast)\tau_D}\big)}{(N-2)a^\ast+a_\ast}\times\frac{e^{-(N-1)^2a^\ast T_D}}{2}\Big)(d_\ast+\varepsilon)
\end{align}
again by Gr\"{o}nwall's inequality and some simple algebra.

Next, applying the estimate of node $i_0$ in step 1 on $i_1$ during time interval $[\hat{t}+\tau_D,t_{k_0}+(N-1)T_D]$, we arrive at
\begin{align}
\sqrt{V_{i_1}(t)}&\leq \frac{ a_\ast\big(1-e^{- ((N-2)a^\ast+a_\ast)\tau_D}\big)}{(N-2)a^\ast+a_\ast}\times\frac{e^{-2(N-1)^2a^\ast T_D}}{2} \theta_{i_0}\nonumber\\
&\ \ \ \ \ \ \ \ \ \ \ +\Big(1-\frac{ a_\ast\big(1-e^{- ((N-2)a^\ast+a_\ast)\tau_D}\big)}{(N-2)a^\ast+a_\ast}\times\frac{e^{-2(N-1)^2a^\ast T_D}}{2}\Big)(d_\ast+\varepsilon)
\end{align}
for all $t\in[t_{k_0}+T_D, t_{k_0}+(N-1)T_D]$.

\noindent{\it Step 3.} Noticing that  the graph is uniformly jointly strongly connected, the analysis of steps 1 and 2 can be repeatedly applied to nodes $i_3,\dots,i_{N-1}$, and eventually we have that for all $i_0,\dots,i_{N-1}$,
\begin{align}
\sqrt{V_{i_m}\big( t_{k_0}+(N-1)T_D\big)}&\leq \Big(\frac{ a_\ast\big(1-e^{- ((N-2)a^\ast+a_\ast)\tau_D}\big)}{(N-2)a^\ast+a_\ast}\Big)^{N-2}\times\frac{e^{-(N-1)^3a^\ast T_D}}{2} \theta_{i_0}\nonumber\\
&\ \ \ \ \ \ \ \ \ \ \ +\Bigg(1-\Big(\frac{ a_\ast\big(1-e^{- ((N-2)a^\ast+a_\ast)\tau_D}\big)}{(N-2)a^\ast+a_\ast}\Big)^{N-2}\times\frac{e^{-(N-1)^3a^\ast T_D}}{2} \Bigg)(d_\ast+\varepsilon)\nonumber\\
&<d_\ast
\end{align}
for sufficiently small $\varepsilon$  because $\theta_{i_0}<d_\ast$ and
$$
\Big(\frac{ a_\ast\big(1-e^{- ((N-2)a^\ast+a_\ast)\tau_D}\big)}{(N-2)a^\ast+a_\ast}\Big)^{N-2}\times\frac{e^{-(N-1)^3a^\ast T_D}}{2} <1
$$
is a constant. This immediately leads to that
\begin{align}
V\big(t_{k_0}+(N-1)T_D\big)<d_\ast,
\end{align}
which contradicts the definition of $d_\ast$.

This completes the proof.
\subsection*{\bf A.3\ \  Proof of Lemma \ref{lemnodeoptimum} }

With Lemma \ref{lemlimit}, we have  that $\lim_{t\rightarrow \infty} V_i(t)=d_\ast^2$ for all $i\in\mathcal{V}$. Thus, $\forall\varepsilon>0, \exists M(\varepsilon)>0$, s.t.,
\begin{align}\label{40}
d_\ast\leq \sqrt{V_i(t)}\leq d_\ast+\varepsilon
\end{align}
for all $i$ and $t\geq M$. If $d_\ast=0$, the desired conclusion follows straightforwardly. Now we suppose $d_\ast>0$.

Assume that there exists a node $i_0$ satisfying $\limsup_{t\rightarrow \infty} \big| x_{i_0}(t)\big|_{\arg\min f_{i_0}}>0$. Then we can find a  time sequence $\{t_k\}_1^\infty$ with $\lim_{k\rightarrow \infty}t_k =\infty$ and a constant $\delta$ such that
\begin{align}\label{42}
\big| x_{i_0}(t_k)\big|_{\arg\min f_{i_0}}\geq\delta, \ k=1,\dots.
\end{align}
Denote also $B_1\doteq\big\{z: |z-z_\ast|\leq d_\ast+1\big \}$ and $G_1=\max\big\{ \nabla f_{i_0}(y):\ y\in B_1\big\}$.
Assumption A1 ensures that $G_1$ is a finite number since $B_1$ is  compact. By taking $\varepsilon=1$ in (\ref{40}),
we see that $x_i(t)\in B_1$ for all $i$ and $t\geq M(1)$.  As a result, we have
\begin{align}\label{41}
\Big|\frac{d}{dt}{x}_{i_0}(t)\Big|=\Big|\sum_{j\in\mathcal{N}_{i_0}(\sigma(t))} a_{i_0 j}(t)(x_j-x_{i_0})+\nabla f_{i_0}(x_{i_0})\Big|\leq 2(n-1) a^\ast (d_\ast+1)+G_1.
\end{align}

Combining  (\ref{42}) and (\ref{41}),   we conclude  that
\begin{align}\label{43}
\big| x_{i_0}(t)\big|_{\arg\min f_{i_0}}\geq \frac{\delta}{2}, \ t\in[t_k,t_k+\tau],
\end{align}
for all $k=1,\dots$, where by definition $\tau=\frac{\delta}{2\big( 2(n-1) a^\ast (d_\ast+1)+G_1\big)}$.

Now we introduce
$$
D_\delta\doteq \min \Big\{f_{i_0}(y)-f_{i_0}(z_\ast):\  \big| x_{i_0}(t)\big|_{\arg\min f_{i_0}}\geq \frac{\delta}{2}\ {\rm and}\ y\in B_1\Big\}.
$$
Then we know $D_\delta >0$ again by the continuity of $f_{i_0}$. According to (\ref{45}), (\ref{40}),  and (\ref{43}), we obtain
\begin{align}
\frac{d}{dt} V_{i_0}(t)
&\leq 2(N-1)a^\ast\Big(-V_{i_0}(t)+(d_\ast+\varepsilon)\sqrt{V_{i_0}(t)}\Big)+f_{i_0}\big(z_\ast\big)-f_{i_0}\big(x_{i_0}(t)\big)\nonumber\\
&\leq   2(N-1)a^\ast (d_\ast+\varepsilon)\varepsilon -D_\delta,
\end{align}
for $t\in [t_k,t_k+\tau]$, $k=1,\dots$. This leads to
\begin{align}\label{46}
V_{i_0}(t_k+\tau)&\leq V_{i_0}(t_k)+\Big(2(N-1)a^\ast (d_\ast+\varepsilon)\varepsilon -D_\delta\Big)\tau \nonumber\\
&\leq d_\ast+\varepsilon+\Big(2(N-1)a^\ast (d_\ast+\varepsilon)\varepsilon -D_\delta\Big)\tau\nonumber\\
&< d_\ast
\end{align}
as long as $\varepsilon$ is sufficiently small so that
$$
\varepsilon \Big(1+2(N-1)a^\ast (d_\ast+\varepsilon)\Big) <D_\delta \tau.
$$
We see that (\ref{46}) contradicts (\ref{40}). The desired conclusion thus follows.

\end{document}